\documentclass[twocolumn,twoside]{article}

\usepackage{a4}
\usepackage{amssymb}
\usepackage{amsmath}
\usepackage{graphicx}
\usepackage{blindtext}

\begin{document}

\title{{\bf Internal Structure of the Jets from Young Stars Simulated at Plasma
Focus Facilities}}

\date{{\normalsize\textit{$^{1}$Lebedev Physical Institute, Russian Academy of Sciences,
Leninskii pr. 53, Moscow, 119991,  Russia\\
$^{2}$Moscow Institute of Physics and Technonogy 
Institutsky per. 9, Dolgoprudny, Moscow obl., 141700,  Russia \\
$^{3}$Keldysh Institute of Applied Mathematics, Russian Academy of Sciences, Miusskaya pl. 4, Moscow, 125047
Russia \\
Received April 15, 2020; revised May 26, 2020; accepted May 26, 2020
}}\\[1ex]
\vspace{0.3cm}
{\small \textit{Pisma v Astronomicheskii Zhurnal} \textbf{46}, No. 7,
pp. 462-472 (2020) 
[in Russian]\\
English translation: \textit{Astronomy Letters}, \textbf{46}, No. 7, pp. 353-370 (2020)}
\\{\small Translated by V. Astakhov}
}

\author{V. S. Beskin$^{1,2}{\footnote{beskin@lpi.ru}}$ \, and I. Yu. Kalashnikov$^{3}$}

\maketitle


{\bf Abstract}--The laboratory simulations of jets from young stars that have been carried out for 
many years at plasma focus facilities allow the internal structure of the active 
regions emerging during the interaction of the jet with the surrounding plasma to 
be studied in detail. We have found a new wide class of solutions for the equations 
of ideal magnetohydrodynamics describing the closed axisymmetric stationary flows 
that are apparently realized in the active regions. Such flows are shown to well 
reproduce the internal structure of the plasma structures observed in laboratory 
simulations of astrophysical jets.

\begin{center}
INTRODUCTION
\end{center}
\noindent

At present, laboratory simulations begin to play an increasingly big role in 
investigating the processes occurring in space. Indeed, despite the fact
that the characteristic lengths and time scales of laboratory experiments are 
smaller than those for real astrophysical sources by many orders of magnitude,
they can be easily scaled for astrophysical situations if both obey the laws 
of ideal magnetohydrodynamics (MHD). This is because the MHD equations have 
no intrinsic scale and, therefore, they can describe both laboratory and 
astrophysical flows (Ryutov et al. 2000).

Transferring the studies of astrophysical objects to a laboratory has a number 
of indubitable advantages. First of all, the parameters of flows can be easily 
varied in a laboratory plasma, which is very important for testing the predictions 
of theoretical models. Next, the time frames of laboratory experiments are small
and, therefore, the dynamics of ongoing processes can be easily followed, whereas 
tracing the dynamics of real astrophysical phenomena can take many decades. 
Furthermore, laboratory experiments can in principle be completely diagnosed, 
while the diagnostics of real astrophysical objects is significantly limited.

One of these directions of laboratory studies is the simulation of astrophysical 
jets. Since nonrelativistic flows are realized in most cases, here one can talk
only about the jets from young stars (Surdin 2001; Bodenheimer 2011). Recall, 
however, that such jets are observed in a great variety of cosmic sources: from 
blazers, active galactic nuclei, and, presumably, gamma-ray bursts to microquasars 
and young stars (see, e.g., Beskin 2005). The jets in these objects have scales 
from megaparsecs (active galactic nuclei) to fractions of a parsec (young stars), 
while the flow velocities range from ultrarelativistic, with a Lorentz factor of 
several tens, to nonrelativistic (for young stars) values. The jets allow an excess 
angular momentum of the ''central engine'' (a black hole, a young star) and the 
accreting matter to be removed in a natural way, which allows, for example, a young 
star to contract to the required sizes. It should also be noted that in almost all 
cases the main energy release occurs in the so-called active regions, where a supersonic 
jet interacts with the ambient medium; in nonrelativistic jets from young stars they 
were first discovered as Herbig-Haro objects (Herbig 1950; Haro 1950).

Now more than six hundred young stars with jets are already known (Arce et al. 2007; 
Ray et al. 2007). Their active regions are bright clumps a few arcsec in size (the 
linear size is $\sim$500-1000 AU) usually surrounded by a bright diffuse envelope. As 
has already been noted, the jet speed exceeds the speed of sound in the jet material. 
Therefore, a shock inevitably appears due to the interaction of a supersonic jet with
the external medium (McKee and Ostriker 2007).

It is clear that the interaction of a jet with interstellar gas has always been the 
focus of attention. The emergence of shocks during the interaction of a supersonic jet 
with the ambient medium was simulated and the role of radiation processes was generally
elucidated already in the 1980–1990s (Norman et al. 1982; Blondin et al. 1990; Stone 
and Norman 1993). Subsequently, to analyze the heating and radiation processes at shocks, 
all of the main ionization and recombination processes were included into consideration 
(Raga et al. 2007). The complex multi-component structure of the ''heads'' was also 
reproduced (Stone and Hardee 2000; Hansen et al. 2017) and the jet-side wind interaction 
was even simulated (Kajdi{\u c} and Raga 2007) (for a review, see also Frank et al. 2014).
Quite a few works on numerical simulations were also associated with an analysis of the 
results obtained at experimental facilities (Ciardi 2010; Bocchi et al. 2013). In all 
numerical experiments the magnetic field actually played a decisive role, allowing the 
main morphological properties of the observed flows to be reproduced.

As regards the laboratory simulations, at present, there are already about ten facilities 
in the world at which the laboratory simulations of astrophysical jets are performed (Ciardi 
et al. 2009; Suzuki-Vidal et al. 2012; Huarte-Espinosa et al. 2012; Albertazzi et al. 2014; 
Belyaev et al. 2018; Bellan 2018; Lebedev et al. 2019; Lavine and You 2019). The jet launch 
was realized both using the $Z$-pinch technology (the MAGPIE facility at the Imperial College,
Great Britain, and the facility at the Cornell University, USA) and through the interaction 
of a super-powerful laser pulse with a target (the LULI-2 facility at the {\' E}cole Polytechnique, France, and the facilities at the University of Rochester, USA, and the Central Research 
Institute for Machine Building, Russia) and at the facilities in which the plasma accelerator
technology was used (the California Institute of Technology and the Washington University,
USA).

One more promising direction of laboratory research on jets is associated with the plasma 
focus technology. These studies were begun several years ago at the National Research Center
''Kurchatov Institute'' at the PF-3 facility (Krauz et al. 2015, 2018; Mitrofanov et al. 
2017) and were then continued at the Institute of Plasma Physics and Laser Fusion (the
PF-1000 facility,Warsaw) and the KPF-4 ''Phoenix'' facility at the Sukhum Physical-Technical 
Institute (Krauz et al. 2017). A large volume of data concerning the internal structure of 
the plasma jet was also accumulated here. This became possible owing to the sufficiently 
large jet sizes, which allow both direct probe measurements of the internal structure of 
the magnetic fields and direct measurements of the plasma flow velocities to be carried 
out. 

Another important feature of the experiments at plasma focus facilities is that the 
plasma outflow moves not in a vacuum, but in the external medium, with this motion being 
supersonic. This fact allows the interaction of real astrophysical jets with interstellar
gas, which, as has already been noted, also occurs in a supersonic regime, to be simulated 
in a laboratory. Furthermore, the possibility to trace the evolution of the plasma outflow 
at distances $\sim$1 m (i.e., greater than its transverse size by tens of times) gives a 
unique opportunity to understand the cause of the stability of jets.

Finally, one more important fact should be emphasized. In contrast to many other laboratory
experiments, not a quasi-stationary cylindrical configuration, but an isolated plasma outflow 
is realized at plasma focus facilities. However, according to astrophysical observations 
(Reipurth et al. 2002; Hansenet al. 2017), nonrelativistic jets from young stars actually
break up into individual fragments (they are all now called Herbig–Haro flows). The 
possibility to directly investigate the structure of such flows is yet another advantage of 
the laboratory studies based on the plasma focus technology. As a result, many of the
questions concerning the stabilizing role of a magnetic field and the gas heating and cooling
dynamics in active regions have been clarified.

\begin{figure}
\begin{center}
\includegraphics[width=220pt]{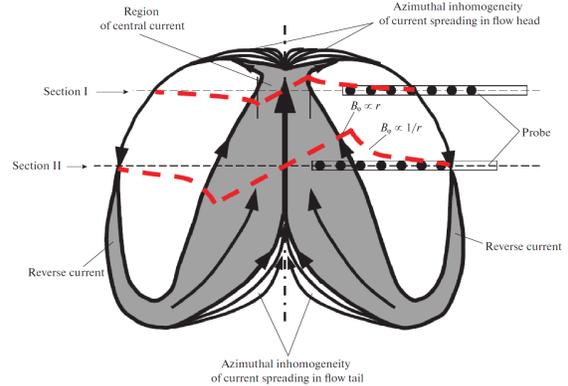}
\caption{The internal structure of the plasma outflow reproduced based on the results 
   obtained at the KPF-4 ''Phoenix'' facility (Krauz et al. 2019). Solid lines 
   indicate the structure of the poloidal magnetic field, the arrows indicate the 
   current circulation scheme; the dashed lines indicate the radial distribution of 
   the toroidal magnetic field in the plasma flow $B_{\varphi}(r)$ in its central 
   region and on the periphery. Two positions of the magnetic probe, I and II, are 
   also shown. 
}
\label{fig01}
\end{center}
\end{figure}

Figure~\ref{fig01} shows the typical shape of the plasma outflow constructed from the 
magnetic probe measurements at the KPF-4 facility (Krauz et al. 2019). In this 
experiment the radial distribution of the toroidal magnetic field was measured 
with a multichannel magnetic probe consisting of coils with a separation between 
the coil centers of 5-6 mm. In this case, the signals from the coils will depend 
on what part of the plasma flow the probe is located in at a given time. For example, 
for section I some of the coils are in the magnetic field of the central current 
outside the zone of its flow and some of them are beyond the zone of the reverse 
current flow, where the magnetic field is zero. In section II it is shown the case 
where some of the coils is in the region of the axial current flow and, accordingly, 
a magnetic field increasing with radius, while the rest are outside the central current, 
in the region of a magnetic field decreasing with radius. An analysis of the signals 
at various times allowed one to determine the radii of the flow of both central
and reverse currents for various experimental conditions and to construct a 
phenomenological plasmoid model.

As we see, the jet is a quasi-toroidal flow at the center of which, according to 
direct probe measurements (dashed line), a 
4
 longitudinal electric current flows. 
The transverse size of the head turns out to be noticeably smaller than that of 
the tail. Of special note is the characteristic funnel in the jet head that
is observed both in laboratory experiments and in real astrophysical sources. 
This study is devoted to explaining this structure (corresponding precisely to
Herbig-Haro flows rather than cylindrical jets). In other words, below we will 
construct a solution of the ideal MHD equations describing a toroidal magnetized
plasma outflow moving in a medium at rest.

Finding a self-consistent configuration of an isolated plasma outflow is also 
important for the numerical simulations of the propagation of laboratory and 
astrophysical jets. The point is that, usually, a continuous inflow into the 
ambient medium from the lower boundary of the computational domain is specified 
as initial conditions in such simulations (Belyaev et al. 2018). In this case, 
the characteristic fragmentary structure of the jets from young stars is produced
by some imposed periodic perturbation of an initially homogeneous flow (Te{\c s}ileanu 
et al. 2012). Since in laboratory conditions, as a rule, we deal with an isolated 
jet, to properly specify the initial conditions, we need to know not only the 
hydrodynamic characteristics of such a jet, but also the magnetic field structure 
whose appropriate choice is a nontrivial problem. Solving the above problem can help 
in choosing suitable initial conditions for the simulations of laboratory jets and, 
as we hope, Herbig-Haro objects.

In the first part of the paper we recall the fundamentals that underlie the method 
of the Grad-Shafranov equation describing stationary axisymmetric flows in the 
approximation of ideal magnetohydrodynamics. In the second part we formulate a new
wide class of solutions for this equation describing closed stationary flows. We 
will restrict ourselves to the case of a subsonic flow, because the interaction of 
the plasma outflow with the external medium of interest to us occurs along the contact 
boundary precisely in this regime. The third part is devoted to simulating the 29

internal structure of the plasma outflow realized at the plasma focus facility. In 
Conclusions we discuss possible astrophysical applications.

\begin{center}
FORMULATION OF THE PROBLEM \\
\vspace{0.2cm}
Basic Equations
\end{center}
First of all, recall the fundamentals of the method of the Grad-Shafranov equation 
that allows axisymmetric stationary configurations to be described in terms of ideal
magnetohydrodynamics in the language of one second-order equation for the magnetic
flux function $\Psi(r, z)$ generally containing five integrals of motion, i.e., five 
quantities conserved on the magnetic surfaces. The full version of this equation
including all five integrals of motion was first formulated by L.S. Soloviev in 1963 
in the third volume of the famous series of collections ''Reviews of Plasma Physics''. 
It is clear that the classical version (Shafranov 1957; Grad 1960), which corresponds 
to static configurations (${\bf v} = 0$) and, therefore, contains only two integrals 
of motion was used in most cases to describe the plasma configurations discussed in
connection with the hot plasma confinement problem (Lao et al. 1981; Atanasiu et al. 
2004; Duez and Mathis 2010). However, the works in which the full version was also 
discussed in connection with a laboratory experiment have appeared in recent years 
(Sonnerup et al. 2004; Guazzotto and Harmeiri 2014; Lopez and Guazzotto 2017). As
regards the astrophysical applications, the full version of the Grad-Shafranov equation 
turned out to be very useful in studying transonic flows in the vicinity of neutron 
stars and black holes (Blandford and Payne 1982; Heyvaerts and Norman 1989; Pelletier
and Pudritz 1992; Beskin 2005). In fact, this direction had been the main method of 
investigating the magnetospheres of compact astrophysical objects for several decades 
until it was ousted by numerical methods.

First of all, let us write the relations defining the electromagnetic fields and the 
velocity of the medium via the integrals of motion:
\begin{eqnarray}
 {\bf B} & = & \frac{{\bf \nabla}\Psi \times {\bf e}_{\varphi}}{2\pi r}
  -\frac{2I}{r c}{\bf e}_{\varphi},
\label{defB}
\\
 {\bf E} & = & -\frac{\Omega_{\rm F}}{2\pi c}{\bf \nabla}\Psi,
\label{defE}
\\
 {\bf v} & = & \frac{\eta_{\rm n}}{\rho}{\bf B}+\Omega_{\rm F}
  r{\bf e}_{\varphi}.
  \label{4a}
\end{eqnarray}
Here, $\rho = m_{\rm p}n$ is the density of the medium, $I$ is the density of the medium{\footnote{The minus sign was introduced in order that the current $I$ be
positive.}}, and $\eta_{\rm n}$ is the ratio of the mass flux to the magnetic flux.
We used the freezing-in condition ${\bf E} + {\bf v} \times {\bf B} /c = 0$ to 
derive Eq. (\ref{4a}).

Owing to the Maxwell equation $\nabla \cdot {\bf B} = 0$ and the continuity equation 
$\nabla \cdot (\rho {\bf v}) = 0$ we obtain
\begin{equation}
\eta_{\rm n} = \eta_{\rm n}(\Psi),
\end{equation}
i.e., $\eta_{\rm n}(\Psi)$ is an integral of motion. The energy flux density (Bernoulli 
integral) $E_{\rm n}$ and angular momentum $L_{\rm n}$ also conserved on the magnetic 
surfaces are written as
\begin{eqnarray}
  E_{\rm n}(\Psi) & = & \frac{\Omega_{\rm F} I}{2\pi c\eta_{\rm n}}
  +\frac{v^2}{2} + w,
  \label{5a}  \\
   L_{\rm n}(\Psi) & = & \frac{I}{2\pi c\eta_{\rm n}}+v_{\varphi}r.
    \label{6a}
\end{eqnarray}
Here, $w$ is the specific enthalpy determined from the thermodynamic relation 
${\rm d}P = \rho{\rm d}w - n T{\rm d}s$. 
The angular velocity $\Omega_{\rm F}(\Psi)$ (magnetic surface equipotentiality
condition) and entropy $s(\Psi)$ will be two more invariants. In what follows, 
we will measure the temperature in energy units; in this case, the entropy $s$ 
is dimensionless.

The definitions introduced above allow the longitudinal current $I$ and toroidal 
velocity $v_{\varphi}$ to be defined as
\begin{eqnarray}
 \frac{I}{2\pi} & = & c\eta_{\rm n}
\frac{L_{\rm n}-\Omega_{\rm F}r^2}{1-{\cal M}^2},
\label{Inrel} \\
   v_{\varphi} & = & \frac{1}{r}\frac{\Omega_{\rm F}r^2
    -L_{\rm n}{\cal M}^2}{1-{\cal M}^2},
    \label{9a}
     \end{eqnarray}
where
\begin{equation}
 {\cal M}^{2}=\frac{4\pi\eta_{\rm n}^{2}}{\rho}
  \label{10a}
   \end{equation}
is the square of the Alfv{\' e}n Mach number (${\cal M}^{2} = v_{\rm p}^2/V_{\rm A, p}^2$,  
where $V_{\rm A, p} = B_{\rm p}/(4 \pi \rho)^{1/2}$ is the Alfv{\' e}n velocity{\footnote{Since 
here we everywhere consider only the axisymmetric configurations, only the poloidal components 
of all vectors play a leading role.}}), and $r$ is the cylindrical coordinate. As regards
the quantity ${\cal M}^2$ itself, it should be determined within the approach considered here 
from the Bernoulli equation (\ref{5a}), which, given the algebraic relations (\ref{Inrel})
and (\ref{9a}), can be written as
\begin{eqnarray}
  \frac{{\cal M}^4}{64\pi^4\eta_{\rm n}^2}\left(\nabla\Psi\right)^2
    = 2r^{2} (E_{\rm n} - w)
    \nonumber  \\
  -\frac{(\Omega_{\rm F}r^2- L_{\rm n}{\cal M}^2)^2}
{(1-{\cal M}^2)^2}
      -2r^2\Omega_{\rm F}\frac{L_{\rm n}-\Omega_{\rm F}r^2}
{1-{\cal M}^2}.
       \label{11a}
        \end{eqnarray}   
Recall that the specific enthalpy $w$ in Eq. (\ref{11a}) should be considered as 
a function of entropy $s$, Mach number ${\cal M}^2$, and integral $\eta_{\rm n}$. 
The corresponding relation is
\begin{equation}
\nabla w
=c_{\rm s}^2 \left(2\frac{\nabla\eta_{\rm n}}{\eta_{\rm n}}
-\frac{\nabla {\cal M}^{2}}{{\cal M}^{2}}\right)
+\left[\frac{1}{\rho}\left(\frac{\partial P}{\partial s}\right)_{n}
+\frac{T}{m_{\rm p}}\right]\nabla s.
\label{12aa}
\end{equation}
In particular, for a polytropic equation of state 
\begin{equation}
P = K(s) \, \rho^{\Gamma},
\label{P}
\end{equation}
when for $\Gamma \neq 1$ we have simply
\begin{equation}
w = \frac{c_{\rm s}^2}{(\Gamma - 1)},
\label{w}
\end{equation}
one can obtain the explicit expression
\begin{equation}
w (s, {\cal M}^2, \eta_{\rm n}) = \frac{\Gamma K(s)}{\Gamma - 1}
\left(\frac{4 \pi \eta_{\rm n}^2}{{\cal M}^2}\right)^{\Gamma - 1}.
\label{wpoly}
\end{equation}
Note that the dependence $K(s)$ should have quite a definite form (see, e.g., 
Zel’dovich et al. 1981):
\begin{equation}
K(s) = K_{0} \, e^{(\Gamma - 1)s},
\label{Ks}
\end{equation}
which will also be used below. As a result, the Bernoulli equation allows the square 
of the Mach number to be expressed via the flux function $\Psi$ and five integrals of 
motion:
\begin{equation}
{\cal M}^{2} = {\cal M}^{2}[\Psi; E_{\rm n}(\Psi), L_{\rm n}(\Psi), 
\Omega_{\rm F}(\Psi),
\eta_{\rm n}(\Psi), s(\Psi)].
\label{12aaM}
\end{equation}

Finally, the force balance condition in a direction perpendicular to the magnetic 
surfaces (we will call it the generalized Grad-Shafranov equation) can be written 
as (Heyvaerts and Norman 1989; Beskin 2005)
\begin{eqnarray}
&&\frac{1}{16 \pi^3 \rho}
\nabla_{k}\left(\frac{1-{\cal M}^2}{r^{2}}
    \nabla^{k}\Psi\right)
+\frac{{\rm d}E_{\rm n}}{{\rm d}\Psi}
\nonumber \\
&&+\frac{\Omega_{\rm F}r^2 - L_{\rm n}}{1-{\cal M}^2} \,
\frac{{\rm d}\Omega_{\rm F}}{{\rm d}\Psi}
+\frac{1}{r^2} \,
\frac{{\cal M}^2 L_{\rm n} - \Omega_{\rm F}r^2}{1-{\cal M}^2}
\frac{{\rm d}L_{\rm n}}{{\rm d}\Psi}
\nonumber \\
&&+\left(2E_{\rm n} - 2 w + \frac{1}{r^2} \,
\frac{\Omega_{\rm F}^2r^4 -2\Omega_{\rm F}L_{\rm n}r^2
+ {\cal M}^2 L_{\rm n}^2}{1-{\cal M}^2}\right) 
\nonumber \\
&&\times \frac{1}{\eta_{\rm n}}\frac{{\rm d}\eta_{\rm n}}{{\rm d}\Psi}
-\frac{T}{m_{\rm p}} \, \frac{{\rm d}s}{{\rm d}\Psi} = 0.
\label{GS}
\end{eqnarray}
Since ${\cal M}^2$, according to (\ref{12aaM}), is now a known function of magnetic flux 
$\Psi$, Eq. (\ref{GS}) is a closed equation that allows the shape of the magnetic surfaces 
to be determined.

We will emphasize once again the main property of the approach considered here that makes 
it most appealing in certain cases. The point is that once Eq. (\ref{GS}) has been solved, 
i.e., once the function $\Psi(r,z)$ (and, hence, the poloidal field structure) has been 
found, all of the remaining quantities can be determined from the algebraic, though implicit,
equations (\ref{Inrel})--(\ref{11a}). Thus, in some cases, it turns out to be possible to 
obtain important information about the properties of flows based on the  analysis of only 
fairly simple algebraic relations without resorting to the solution of  the nonlinear 
differential equation (\ref{GS}).

\begin{center}
Grad-Shafranov Equation
\vspace{0.2cm}
\end{center}
Recall now how the transition from Eq. (\ref{GS}) to the Grad-Shafranov equation, i.e., 
to the equation describing static configurations (${\bf v} = 0$), occurs within the 
general approach. For this purpose, let us first set $\Omega_{\rm F} = 0$ and 
$\eta_{\rm n} = $ const in Eq. (\ref{GS}). This already allows us to get rid of two 
fairly cumbersome terms. Next, we pass to the limits $\eta_{\rm n} \rightarrow 0$  
(i.e., ${\cal M}^2 \rightarrow 0$) and $L_{\rm n}\rightarrow \infty$, so that
\begin{equation}
I(\Psi) = 2 \pi c \eta_{\rm n}L_{\rm n}(\Psi) = O(1).    
\end{equation} 
In this case, the Bernoulli integral will be written simply as $E_{\rm n} = w$.    

Multiplying now Eq. (\ref{GS}) by $16 \pi^3 r^2 \rho$ while expanding the product 
${\cal M}^2L_{\rm n}$ as  $4\pi \eta_{\rm n}^2 L_{\rm n}/\rho$ and using the 
thermodynamic relation ${\rm d}P = \rho{\rm d}w - n T {\rm d}s$, we
finally obtain in cylindrical coordinates $(r, \varphi, z)$
\begin{equation}
\Psi_{rr} - \frac{\Psi_{r}}{r} + \Psi_{zz} 
+ 16 \pi^2 I\frac{{\rm d}I}{{\rm d}\Psi} 
+ 16 \pi^3 r^2 \frac{{\rm d}P}{{\rm d}\Psi} = 0.
\label{GSGS}
\end{equation}
As we see, the Grad–Shafranov equation requires specifying only two integrals 
of motion, $I(\Psi)$ and $P(\Psi)$. Now there is already no need to add the
Bernoulli equation (which turns out to hold identically).

Clearly, the Grad-Shafranov equation (\ref{GS}) has been studied reasonably well 
(Landau and Lifshitz 1982) and for the simplest linear dependences
\begin{eqnarray}
I(\Psi) & = & a\Psi,
\nonumber \\
P(\Psi) & = & b\Psi + P_{0},
\end{eqnarray}
when it becomes linear,
\begin{equation}
\Psi_{rr} - \frac{\Psi_{r}}{r} + \Psi_{zz} 
+ 16 \pi^2 a^2  \Psi  + 16 \pi^3 br^2 = 0,
\label{GSGSGS}
\end{equation}
analytical solutions were obtained. In particular, the cylindrical solution of Eq. 
(\ref{GSGS}) at $P(\Psi) =$ const
\begin{equation}
\Psi(r) = k r J_{1}(kr),
\label{k}
\end{equation}
which leads to a classical dependence of the fields $B_{\varphi}$ and
$B_{z}$ on $r$,
\begin{eqnarray}
B_{\varphi}(r) & = & B_{0} J_{1}(k r), 
\nonumber \\
B_{z}(r) & = & B_{0} J_{0}(kr),
\end{eqnarray}
is well known. Here, $J_{0}(x)$ и $J_{1}(x)$ are Bessel functions and we set 
$k  = 4 \pi a$.
Below we will use the obvious two-dimensional generalization of this solution
\begin{equation}
\Psi(r,z) = C k_{1} r J_{1}(k_{1} r) \cos(k_{2}z + \phi_{0})
- \frac{\pi b}{a^2} r^2,
\label{solution}
\end{equation}
where $C$ and $\phi_{0}$ are arbitrary constants. As is easy to verify, 
(\ref{solution}) is indeed a solution of (\ref{GSGSGS}) when the condition
\begin{equation}
\sqrt{k_{1}^2 + k_{2}^2} = 4 \pi a
\label{k1k2}
\end{equation}
is fulfilled.

\begin{center}
NEW CLASS OF SOLUTIONS FOR NONZERO VELOCITY
\end{center}
Unfortunately, it should be immediately noted that the solution considered above
--- here, of course, we are dealing with some basis in terms of which any solution 
of Eq.(\ref{GSGSGS}) can be expanded --- cannot be used to analyze the internal 
structure of a plasma outflow propagating in an external medium. This is because the 
magnetic surfaces are isobaric ($P$ = const) and, therefore, this solution (in 
the jet rest frame) cannot be joined to the external flow in which the pressure
along the boundary is not constant.

However, let us show that the family of solutions for Eq. (\ref{solution}) 
considered above has a much wider range of applicability. It turns out that 
this family remains a basis even for the more complex problem in which all
five integrals are nonzero. To show this, let us again multiply Eq. (\ref{GS})
by $16 \pi^3 r^2 \rho$and consider the limit ${\cal M}^2 \ll 1$, corresponding 
to a subsonic flow. After a
rearrangement of terms, we then obtain
\begin{eqnarray}
&&r^2 \,\nabla_{k}\left(\frac{1}{r^{2}} \nabla^{k}\Psi\right)
\nonumber \\
&& + 16 \pi^3 \rho {\cal M}^2
\left(L_{\rm n}\frac{{\rm d}L_{\rm n}}{{\rm d}\Psi}
+  L_{\rm n}^2 \frac{1}{\eta_{\rm n}}\frac{{\rm d}\eta_{\rm n}}{{\rm d}\Psi}
\right)
\nonumber \\
&&+16 \pi^3 r^2 \rho \left(\frac{{\rm d}E_{\rm n}}{{\rm d}\Psi} 
+2 E_{\rm n} \frac{1}{\eta_{\rm n}}\frac{{\rm d}\eta_{\rm n}}{{\rm d}\Psi}
- \Omega_{\rm F} \frac{{\rm d}L_{\rm n}}{{\rm d}\Psi}\right.
\nonumber \\
&&\left.- L_{\rm n} \frac{{\rm d}\Omega_{\rm F}}{{\rm d}\Psi} 
 - 2\Omega_{\rm F}L_{\rm n}
\frac{1}{\eta_{\rm n}}\frac{{\rm d}\eta_{\rm n}}{{\rm d}\Psi} \right)
\nonumber \\
&& + 16 \pi^3 r^4 \rho \left(\Omega_{\rm F} \frac{{\rm d}\Omega_{\rm F}}{{\rm d}\Psi} 
+\Omega_{\rm F}^2\frac{1}{\eta_{\rm n}}\frac{{\rm d}\eta_{\rm n}}{{\rm d}\Psi}\right)
\nonumber \\
&&-16 \pi^3 r^2 \rho\left(
2 w \frac{1}{\eta_{\rm n}}\frac{{\rm d}\eta_{\rm n}}{{\rm d}\Psi}
+\frac{T}{m_{\rm p}} \, \frac{{\rm d}s}{{\rm d}\Psi}\right) = 0.
\label{GS1}
\end{eqnarray}

Since the coefficient in front of the parenthesis in the second term does not 
contain explicitly the density \mbox{$\rho = \rho({\cal M}^2, \Psi)$} due to condition
(\ref{10a}), it may be retained in the Grad-Shafranov equation provided that the 
second term is linear by $\Psi$. As regards the remaining terms, for the equation 
to be linear, all of them must be set equal to zero. As a result, we obtain the 
following general relations between the integrals of motion whereby the generalized 
Grad-Shafranov equation is linear:
\begin{eqnarray}
\Omega_{\rm F}(\Psi) & = & \frac{\Omega_{0}}{\eta_{\rm n}(\Psi)};
\qquad \qquad \qquad \Omega_{0} = {\rm const}, 
\label{Omn}\\
L_{\rm n}(\Psi) & = & \frac{A}{\eta_{\rm n}(\Psi)} \Psi 
+  \frac{C}{\eta_{\rm n}(\Psi)};
\quad  A, C = {\rm const}, 
\label{Ln}\\
E_{\rm n}(\Psi) & = & \frac{E_{0}}{\eta_{\rm n}^2(\Psi)} 
+ \Omega_{\rm F}(\Psi)L_{\rm n}(\Psi);  \, E_{0} = {\rm const}, 
\label{En} \\
s(\Psi) & = & s_{0} - 2 C_{\rm p} \ln \eta_{\rm n}(\Psi),
\qquad  s_{0} = {\rm const}.
\label{sn}
\end{eqnarray}
In the last equation we took into account the above mentioned thermodynamic 
relations for a polytropic equation of state, for which the heat capacity 
$C_{\rm p} = \Gamma/(\Gamma - 1)$. In what follows, we will always set $C = 0$, 
because the angular momentum $L_{\rm n}$ must be zero on the rotation axis
($\Psi = 0$): $L_{\rm n}(0) = 0$.

As a result, the Grad-Shafranov equation will again be written as 
\begin{equation}
\frac{\partial^2 \Psi}{\partial r^2} - \frac 1 r \frac{\partial\Psi}{\partial r} + 
\frac{\partial^2 \Psi}{\partial z^2} + 64\pi^4 A^2 \Psi =0,
\label{GSGS'}
\end{equation}
i.e., the basis of its solution will not change. In turn, the Bernoulli equation 
defining the square of the Mach number ${\cal M}^2$ (along with all the remaining
flow parameters) takes the form (cf. Guazzotto and Harmeiri 2014)
\begin{eqnarray}
{\cal M}^4\left[\frac{(\nabla\Psi)^2}{64\pi^4\eta_{\rm n}^2r^2} 
+\frac{L_{\rm n}^2}{r^2}\right]
= 2(E_{\rm n} - \Omega_{\rm F}L_{\rm n}) 
\nonumber \\
- 2 w({\cal M}^2, \eta_{\rm n}, s)
+r^2\Omega_{\rm F}^2,
\label{Ber}
\end{eqnarray}
where the specific enthalpy $w({\cal M}^2, \eta_{\rm n}, s)$ is specified
by relation (\ref{wpoly}). When deriving (\ref{Ber}),  we again passed to 
the limit ${\cal M}^2 \ll 1$ wherever possible.

We conclude this section by noting yet another interesting fact. If the integrals 
of motion are chosen in the form 
\begin{eqnarray}
\eta_{\rm n}(\Psi) & = & \eta_0 \, e^{\sigma\Psi},\\
\Omega_{\rm F}(\Psi) & = & \Omega_0 \, e^{-\sigma\Psi},\\
L_{\rm n}(\Psi) & = & \frac{A}{\eta_{0}} \Psi \, e^{-\sigma\Psi},\\
E_{\rm n}(\Psi) & = & E_0 \, e^{-2\sigma\Psi} + \Omega_{\rm F}(\Psi) L_{\rm n}(\Psi),\\
s(\Psi) & = & s_0 - 2C_{\rm p}\sigma\Psi,
\end{eqnarray}
where $\sigma$ can have any sign, then all terms in the Bernoulli equation will 
contain the factor $e^{-2\sigma\Psi}$. This follows both from the form of the 
integrals itself and from the explicit expression (\ref{wpoly}) for the specific 
enthalpy $w$ and the condition $K(s) = K_{0} \,e^{(\Gamma - 1) s}$ (\ref{Ks}). 
As a result, we have after cancelations
\begin{eqnarray}
{\cal M}^4\left[\frac{(\nabla\Psi)^2}{64\pi^4\eta_{0}^2r^2} 
+\frac{A^2\Psi^2}{\eta_{0}^2 r^2}\right]
= 2E_{0} 
\nonumber \\
- 2 \frac{\Gamma K_{0}(4 \pi \eta_{0}^2)^{\Gamma - 1}}{(\Gamma -1){(\cal M}^2)^{\Gamma - 1}} 
+ r^2\Omega_{0}^2.
\label{Bern}
\end{eqnarray}
Accordingly, for the temperature $T$ in this case we obtain 
$T = T_{0} \, e^{-2\Gamma\sigma\Psi}$.
 
One more very important remark should be made here. As is well known 
(Heyvaerts and Norman 1989; Sonnerup et al. 2004), caution should be 
exercised when discarding the small summand ${\cal M}^2$ in the first 
term of the Grad-Shafranov equation (\ref{GS}) because it is due to 
this summand that the Grad-Shafranov equation becomes hyperbolic even 
in the region ${\cal M}^2 < 1$, or, more specifically, in the region
where the poloidal velocity $v_{\rm p}$ lies within the range
$V_{\rm cusp, p} < v_{\rm p} < c_{\rm s}$. Here
\begin{equation}
V_{\rm cusp, p} = \frac{c_{\rm s}V_{\rm A,p}}{(c_{\rm s}^2 + V_{\rm A}^2)^{1/2}}
\label{Vcusp}
\end{equation}
is the so-called cusp velocity and we consider the case of $c_{\rm s} < V_{\rm A}$ 
here. Therefore, the condition 
\begin{equation}
v_{\rm p} \ll V_{\rm cusp, p}
\end{equation}
should also be added to the applicability condition for the elliptic equation 
(\ref{GSGS}) ${\cal M}^2 \ll 1$.

Qualitatively, however, the conditions under which the approximation considered 
by us remains valid can be derived directly from relation (\ref{Bern}). Indeed, 
since Eq. (\ref{GSGS}) is linear, the potential $\Psi$ (along with the magnetic 
field) can be made arbitrarily large. On the other hand, according to the Bernoulli 
equation (\ref{Bern}), the square of the Mach number ${\cal M}^2$ is inversely
proportional to $\Psi$, so that the Mach number can always be made arbitrarily 
small for sufficiently strong magnetic fields. Accordingly, the cusp velocity 
also becomes large for a sufficiently strong magnetic field.

\vspace{0.5cm}
\begin{center}
RESULTS
\end{center}
Now we can turn to our main goal --- constructing the solution that describes 
the internal structure of the plasma outflow realized at the KPF-4 ''Phoenix''
facility. For this purpose, it is natural to pass to the reference frame in 
which the plasma outflow is at rest. The problem is then reduced to determining 
the shape of the contact discontinuity separating the plasma outflow and the 
plasma inflow at which the condition for the total pressures being equal is 
fulfilled. The solution in the inner region is reduced to finding the coefficients
$C_{k}$ and $\phi_{k}$ in the expansion
\begin{equation}
\Psi(r,z) = \sum_{k} C_{k} k r J_{1}(k r)\cos(k_{2}z + \phi_{k}),
\label{sol}
\end{equation}
where now
\begin{equation}
k_{2} = \sqrt{64 \pi^4 A^2 - k^2}.
\label{k'}
\end{equation}

\begin{figure}
	\centering
	\includegraphics[width=0.9\linewidth]{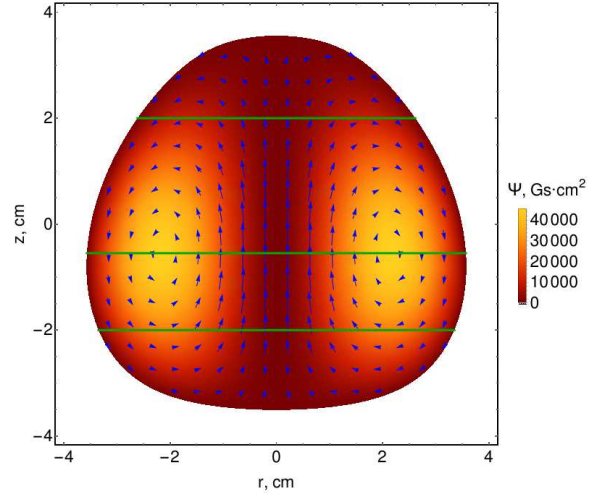}
	\caption{Flow structure within the plasma outflow. The arrows indicate the 
	velocities ${\bf v}$. The poloidal magnetic field ${\bf B}$ points in the same 
	direction, while the electric current density ${\bf j}$ points in the opposite 
	direction. The color represents the potential $\Psi(r, z)$. The three sections 
	used in the succeeding figures are also shown.}
	\label{fig:Psi}
\end{figure}

\begin{figure*}[!ht]
	\begin{minipage}{0.32\linewidth}
		\center{\includegraphics[width=1\linewidth]{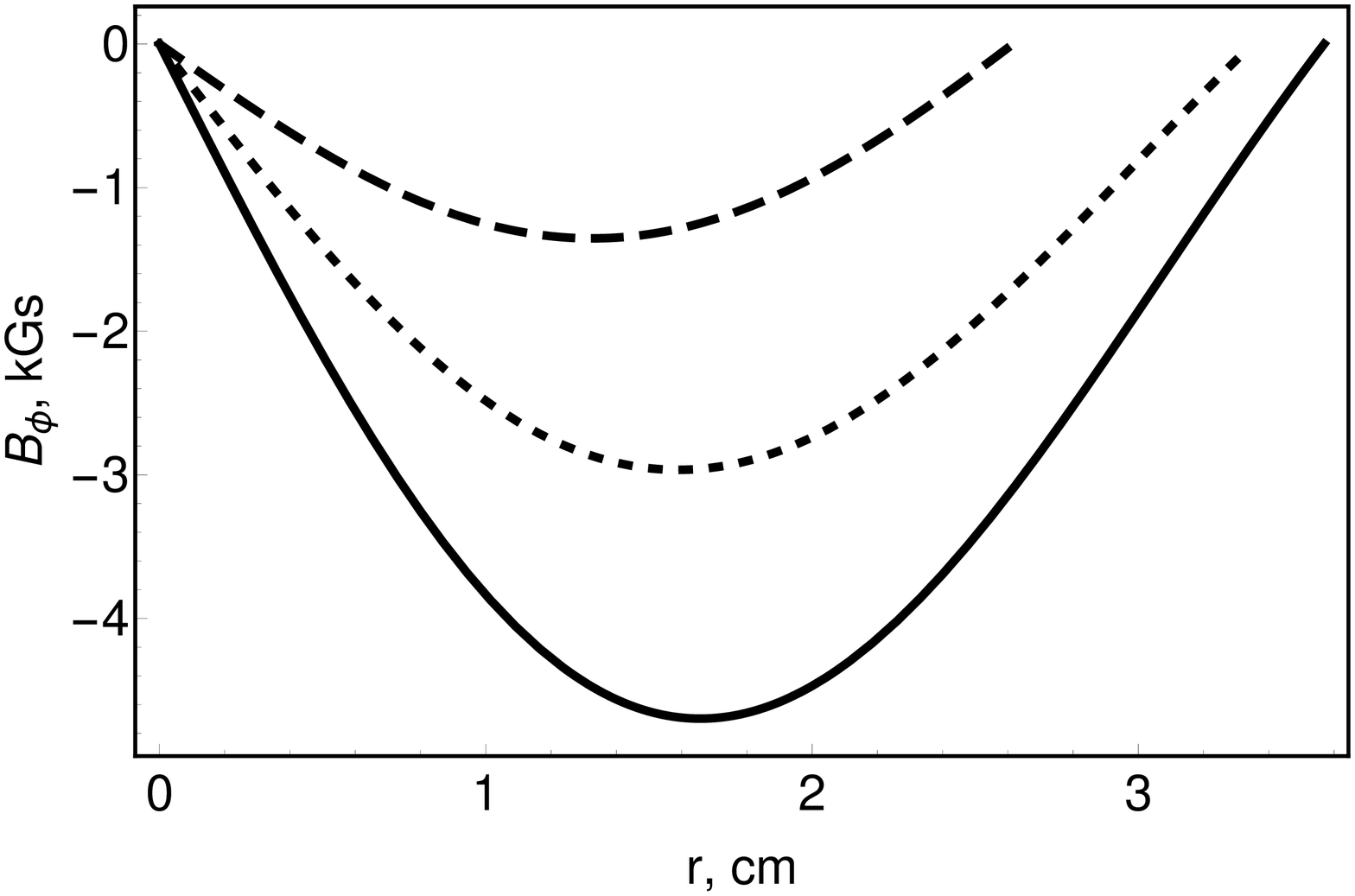} }
	\end{minipage}
	\hfill
	\begin{minipage}{0.32\linewidth}
		\center{\includegraphics[width=1\linewidth]{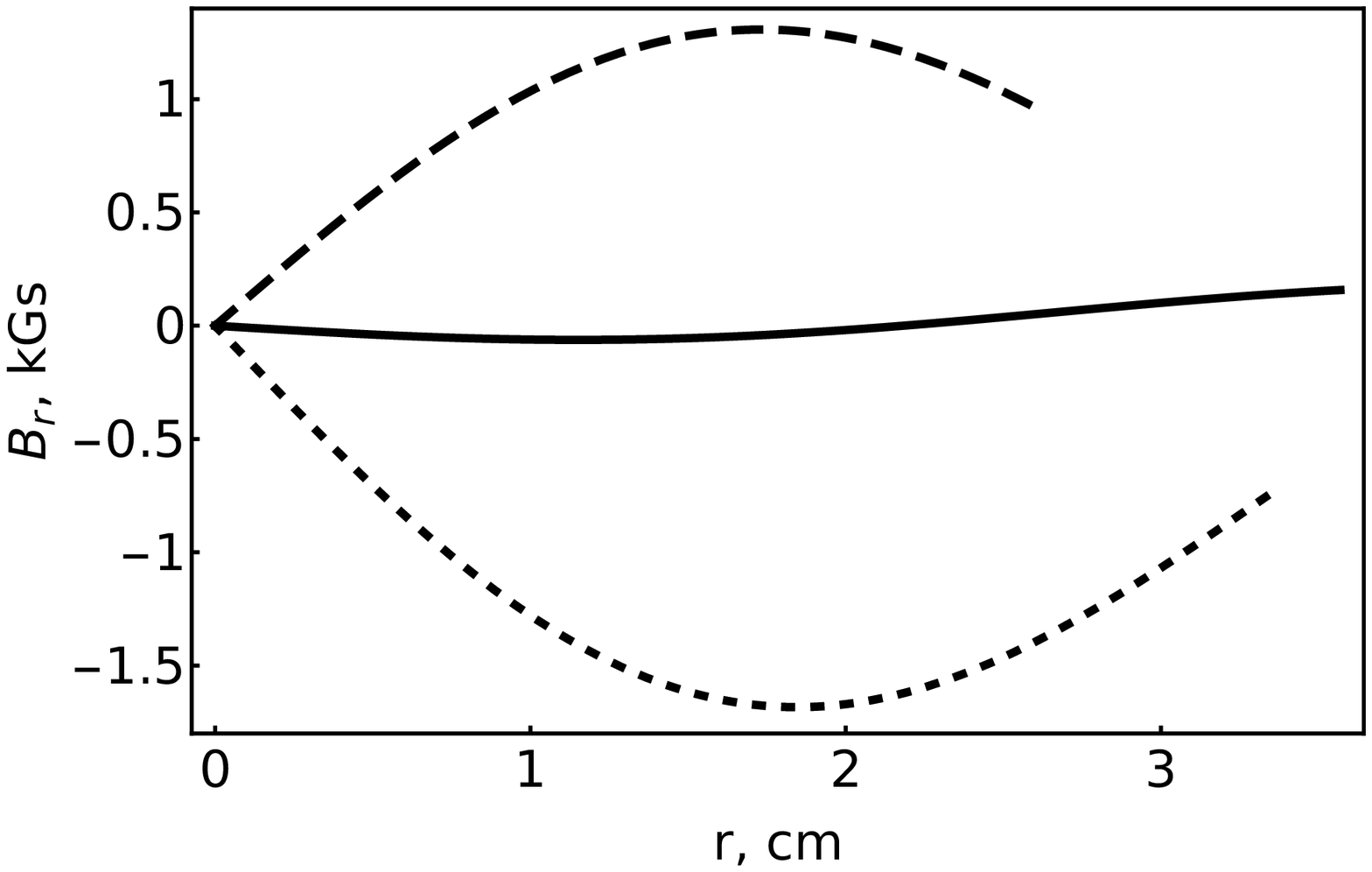}  }
	\end{minipage}
	\hfill
	\begin{minipage}{0.32\linewidth}
		\center{\includegraphics[width=1\linewidth]{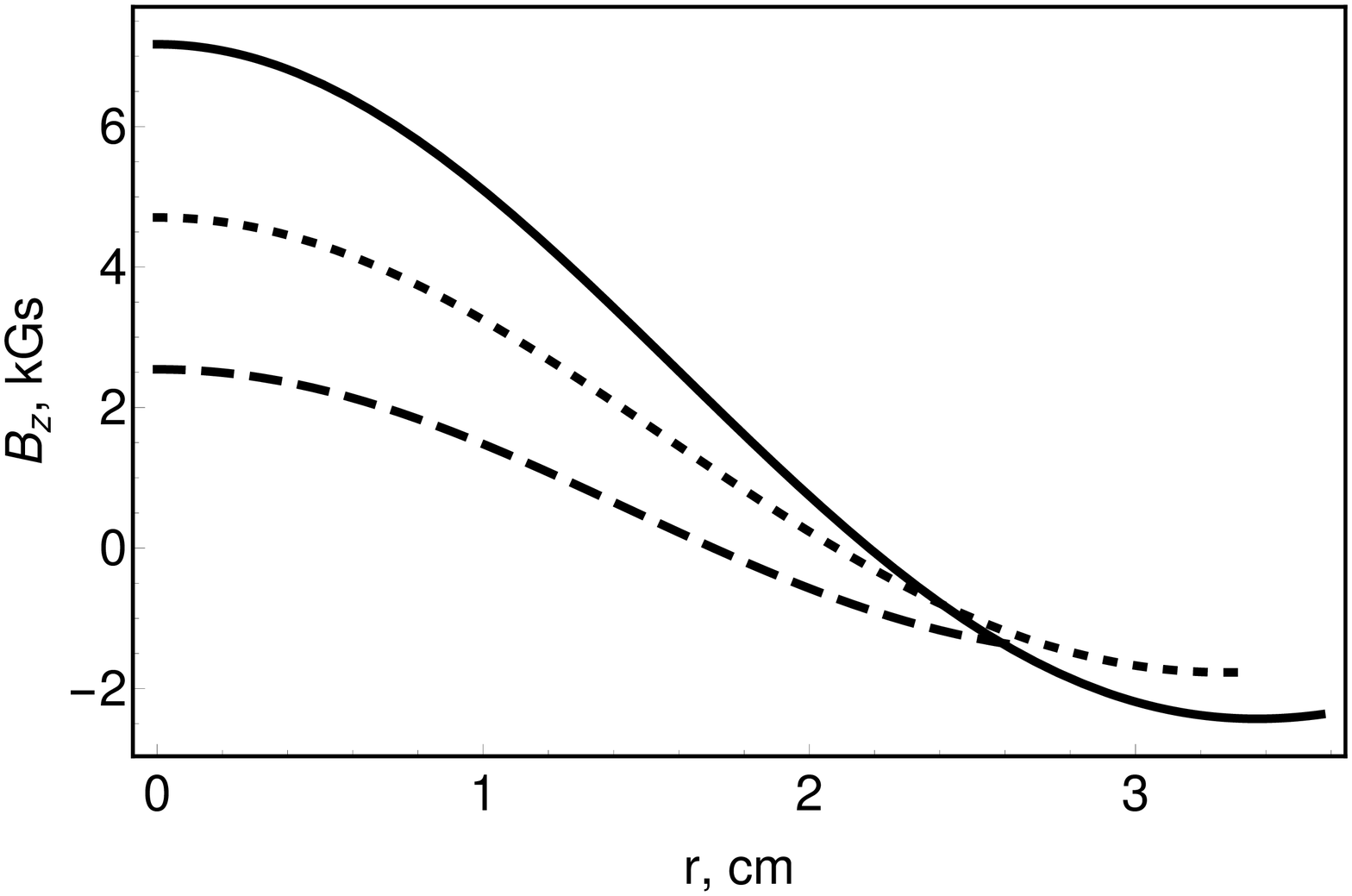} }
	\end{minipage}
	\vfill
	\begin{minipage}{0.32\linewidth}
		\center{\includegraphics[width=1\linewidth]{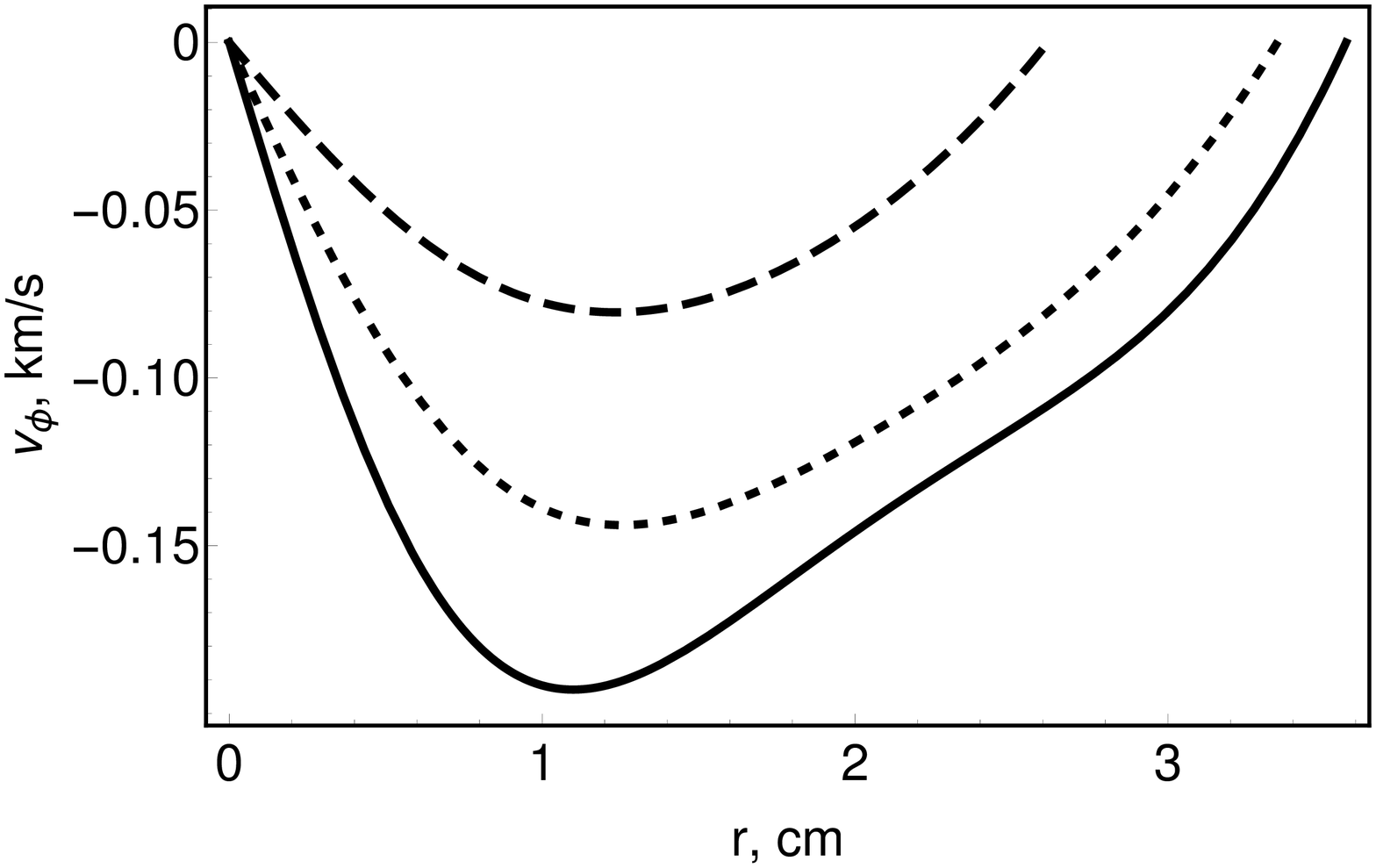}  }
	\end{minipage}
	\hfill
	\begin{minipage}{0.32\linewidth}
		\center{\includegraphics[width=1\linewidth]{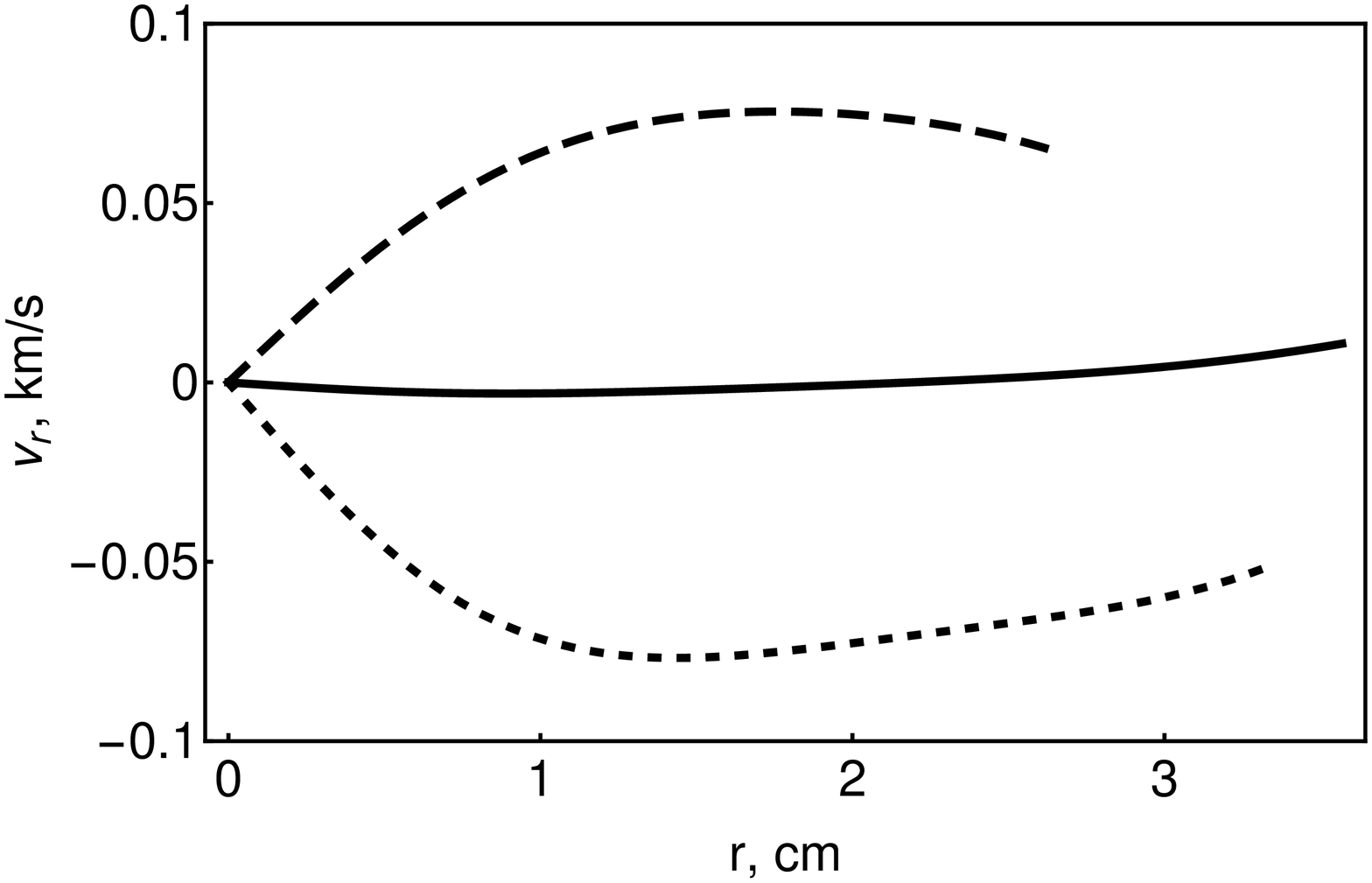}  }
	\end{minipage}
	\hfill
	\begin{minipage}{0.32\linewidth}
		\center{\includegraphics[width=1\linewidth]{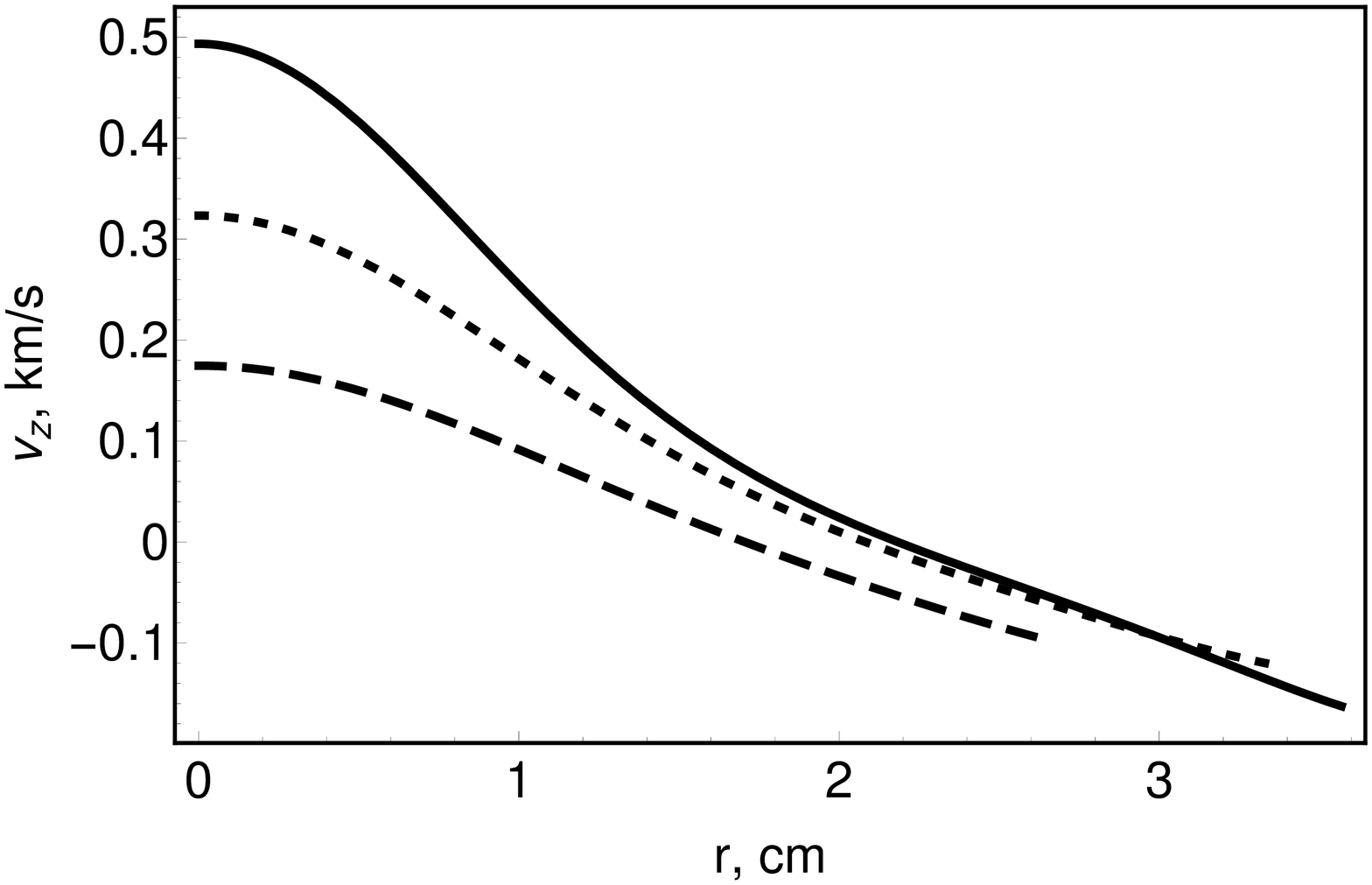}  }
	\end{minipage}
	\caption{Radial distributions of the magnetic field (upper row) 
	and velocity (lower row) at various heights: the solid, dashed,
    and dotted lines correspond to $z = -0.55$ (the middle line in 
    Fig.~\ref{fig:Psi} on which the maximum density is reached), 
    2, and $-2$ cm, respectively.}
\label{fields}	
\end{figure*}

\begin{figure*}[!ht]
	\begin{minipage}{0.32\linewidth}
		\center{\includegraphics[width=1\linewidth]{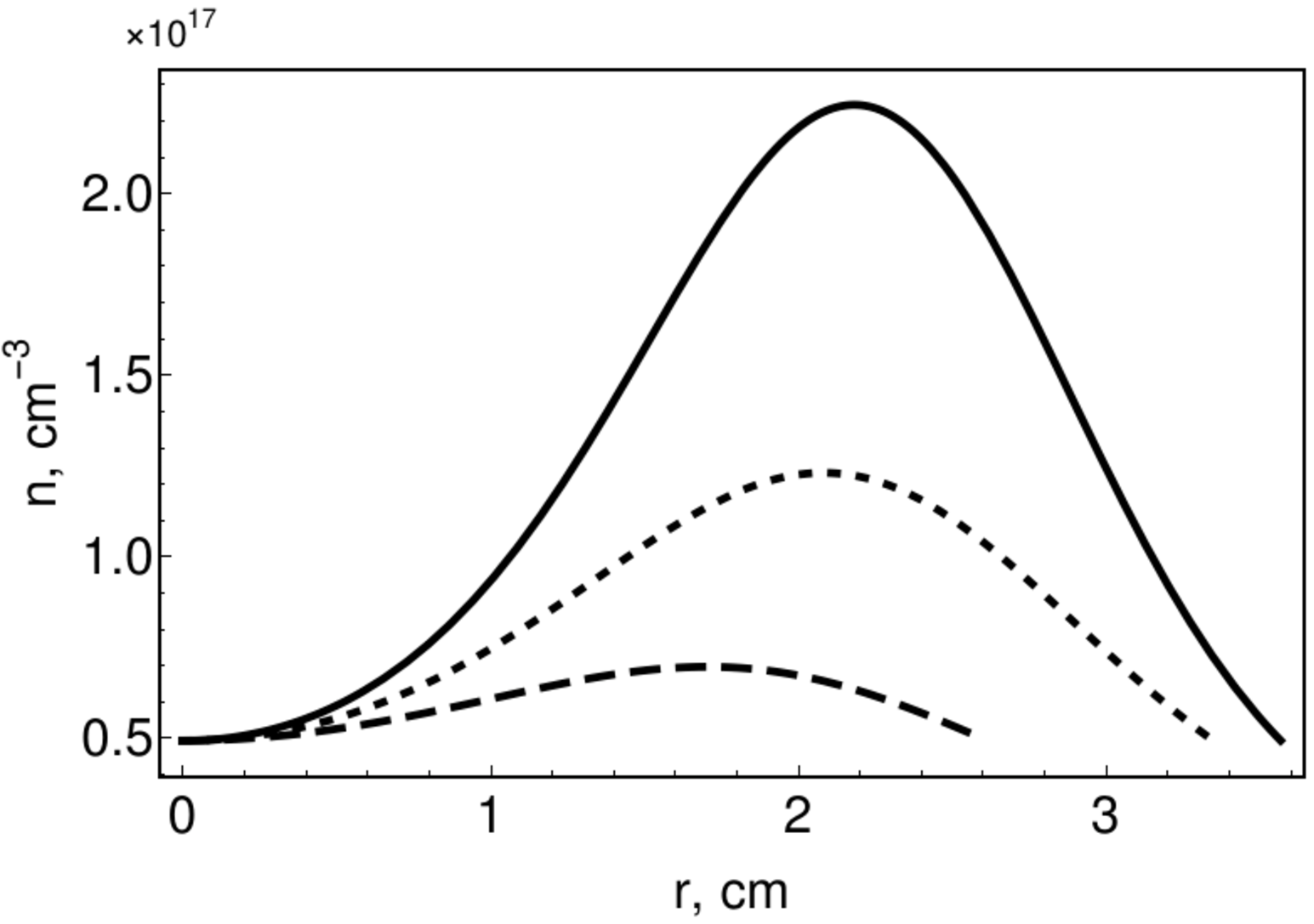} }
	\end{minipage}
	\hfill
	\begin{minipage}{0.32\linewidth}
		\center{\includegraphics[width=1\linewidth]{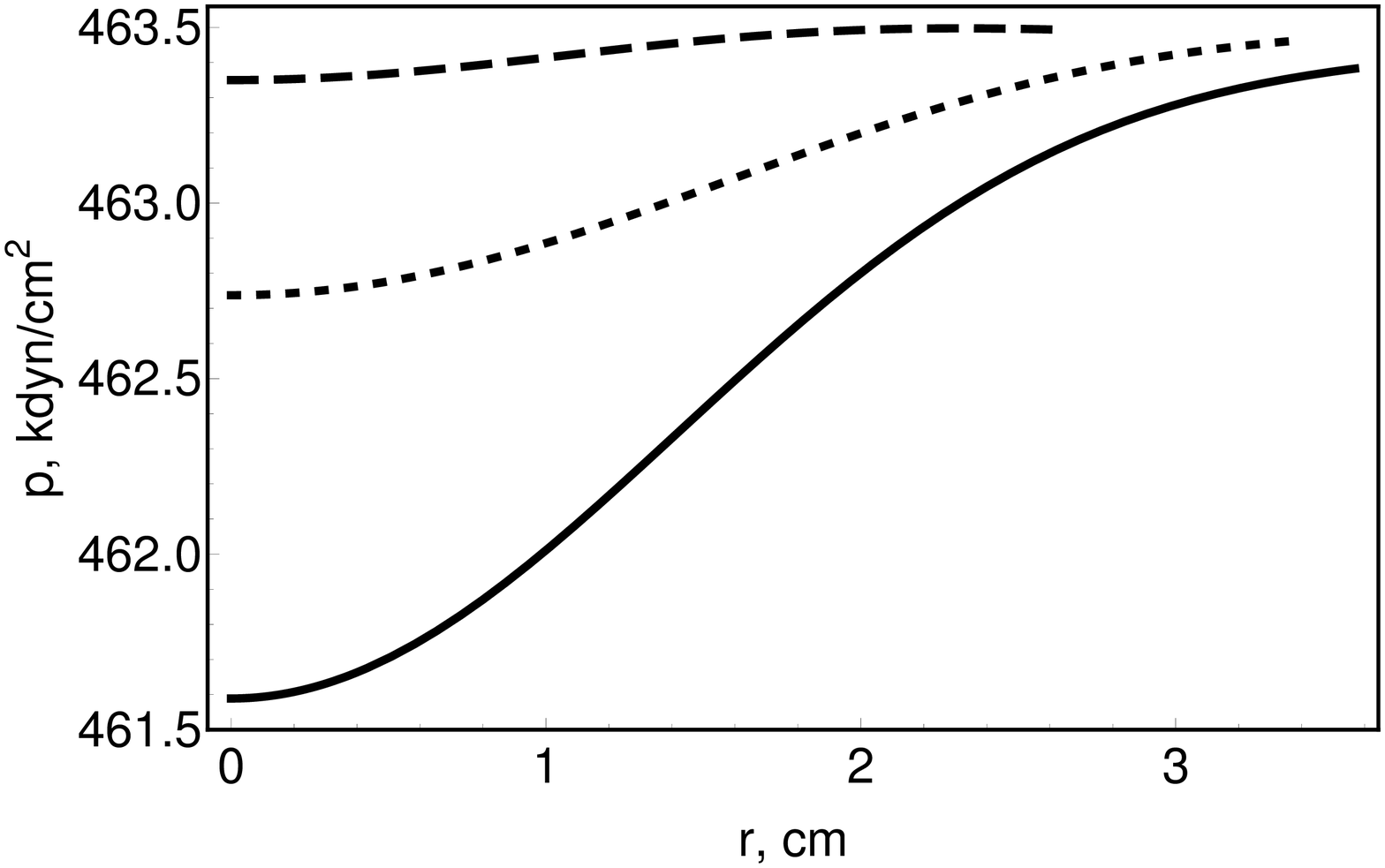}  }
	\end{minipage}
	\hfill
	\begin{minipage}{0.32\linewidth}
		\center{\includegraphics[width=1\linewidth]{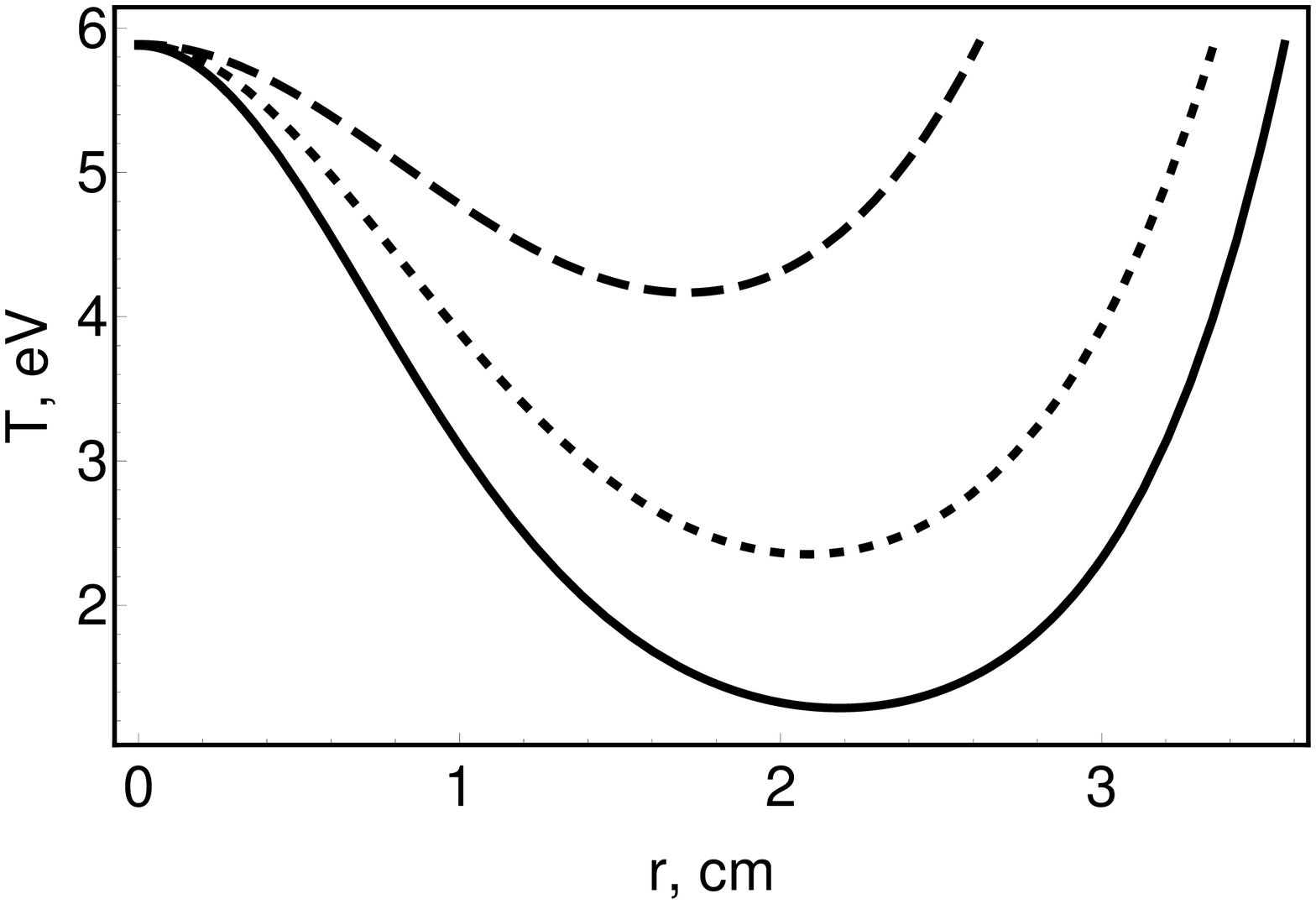} }
	\end{minipage}
	\caption{Radial distributions of the number density, pressure, and 
	temperature at various heights: the solid, dashed, and dotted lines 
	correspond to $z = -0.55$ (the height of the density maximum), 2, 
	and $-2$ cm, respectively.}
\label{density}	
\end{figure*}

\begin{table}
\caption{Parameters of the solutions (\ref{solution}) of the linear
equation (\ref{GSGSGS}) for $\Psi_0=2.4 \times 10^4$ G cm$^{2}$.}
\vspace{0.3cm}
\centering
\begin{tabular}{c|ccc|ccc|ccc|ccc|c}
\hline
 &&&&&&&&&&&&& \\
  $k$      &&  2.1  &&&  0.9  &&&  0.8  &&& 0.2  &&  0.1   \\
 &&&&&&&&&&&&& \\
\hline
 &&&&&&&&&&&&& \\
$\phi_0$   &&  0.0  &&&  0.0  &&&  1.0  &&&  1.2  &&  1.7   \\ 
 &&&&&&&&&&&&& \\
$\frac{C}{\Psi_0}$ &&  1.0  &&&  0.3  &&&  0.4  &&& 0.5  && 0.6    \\
 &&&&&&&&&&&&& \\
\hline
\end{tabular}
\label{table1}
\end{table}

When constructing the solution, we chose the input parameters so that 
they corresponded most closely to the laboratory experiment. Therefore, 
the external medium was simulated by a homogeneous hydrodynamic flow 
(particle number density $n_{\rm e} = 2 \cdot 10^{16}$ cm$^{-3}$, 
velocity $v_{z} = - 100$ km/s). The remaining parameters were chosen 
so that the transverse size of the jet,  as in the laboratory simulations, 
was a few centimeters. Finally, for a complete closure of the current 
at the outer boundary of the plasma outflow, according to (\ref{Inrel}), 
we set $\Omega_0=0$. Note that due to this condition, according to 
(\ref{Inrel}), in the limit ${\cal M}^2 \ll 1$ we obtail 
\begin{equation}
I = 2 \pi c \eta_{\rm n}(\Psi)L_{\rm n}(\Psi),
\label{IPsi}
\end{equation}
so that the electric current again turns out to be an integral of motion. 
This implies that the electric current ${\bf j}_{\rm p}$ will flow along 
magnetic field lines. Due to the minus sign in Eq. (\ref{defB}), its 
direction will be opposite to the direction of the magnetic field.

As a result, it turned out that to determine the structure of the plasma outflow 
at the boundary of which the condition for the balance of the total pressures
with an inflow is fulfilled, we may restrict ourselves with a good accuracy only 
to five solutions (\ref{solution}) of the linear equation (\ref{GSGSGS}). Their 
parameters are given in Table~\ref{table1}. They correspond to the following 
quantities defining the integrals of motion (all quantities are in cgs):
$A = 0.016$ cm$^{-1}$, \mbox{$\eta_0=1 \times 10^{-5}$ g cm$^{-2}$ s$^{-1}$ G$^{-1}$,}   
\mbox{$E_0 = 7\times 10^{11}$ cm$^{2}$ s$^{-2}$,} and $K_0=2 \times 10^{15}$. 
It is convenient to represent the quantity $\sigma$ in fractions of $\Psi_0$: 
$\sigma=0.4/\Psi_0$. Finally, the polytropic index was chosen as for a monoatomic 
gas: $\Gamma=5/3$.

Figure~\ref{fig:Psi} shows the shape of the plasma outflow and the distribution 
of flow velocities ${\bf v}$ within the jet. According to relations (\ref{4a}) and 
(\ref{IPsi}), the magnetic field ${\bf B}$ points in the same direction, while the 
electric current density ${\bf j}$ points in the opposite direction. The color 
represents the potential $\Psi(r, z)$. The three sections that are used in the 
succeeding figures are also shown.

We see that the solution found by us does reproduce well the main morphological 
characteristics --- the increase in the width of the plasma outflow in its tail
and the presence of a characteristic funnel in its head. As shown in Fig.~\ref{fields}, 
the transverse distribution of the toroidal magnetic field Bϕ is also well 
reproduced for the specified choice of integrals (the left panel in the upper 
row). Here, different curves correspond to different heights: the solid, dashed, 
and dotted lines correspond to $z = -0.55$ (the middle section in
Fig.~\ref{fig:Psi} on which the potential Ψ and density reach their maxima), 2, 
and $-2$ cm, respectively. As regards the remaining magnetic field components and the
structure of the flow itself, their comparison with the experimental data is yet 
to be made in future.

\begin{figure}[!ht]
	\center{\includegraphics[width=0.8\linewidth]{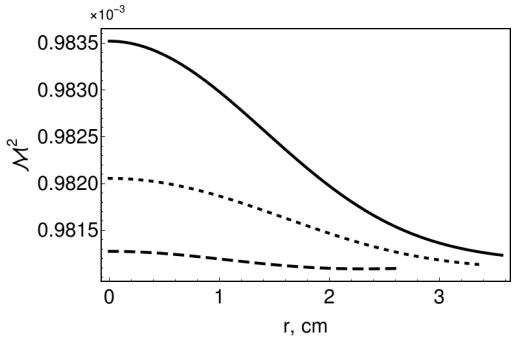} }
	\caption{Mach number ${\cal M}^2$ at various heights: the solid, dashed,
	 and dotted lines correspond to $z = -0.55$, 2, and $-2$ cm,
respectively.}	
\label{Mach}
\end{figure}

\begin{figure*}[!ht]
	\begin{minipage}{0.32\linewidth}
		\center{\includegraphics[width=1\linewidth]{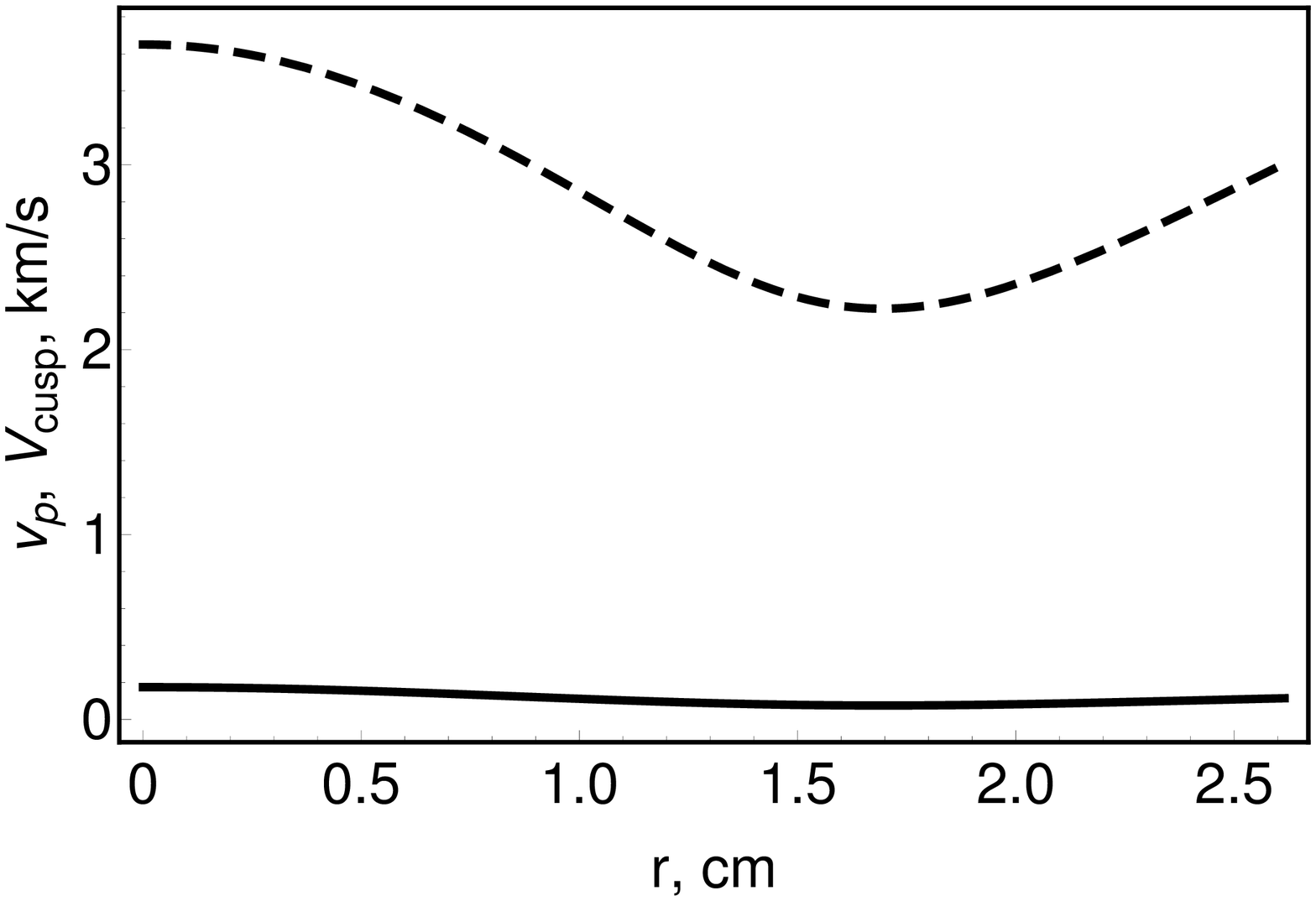} } \\ {\small a)}
	\end{minipage}
	\hfill
	\begin{minipage}{0.32\linewidth}
		\center{\includegraphics[width=1\linewidth]{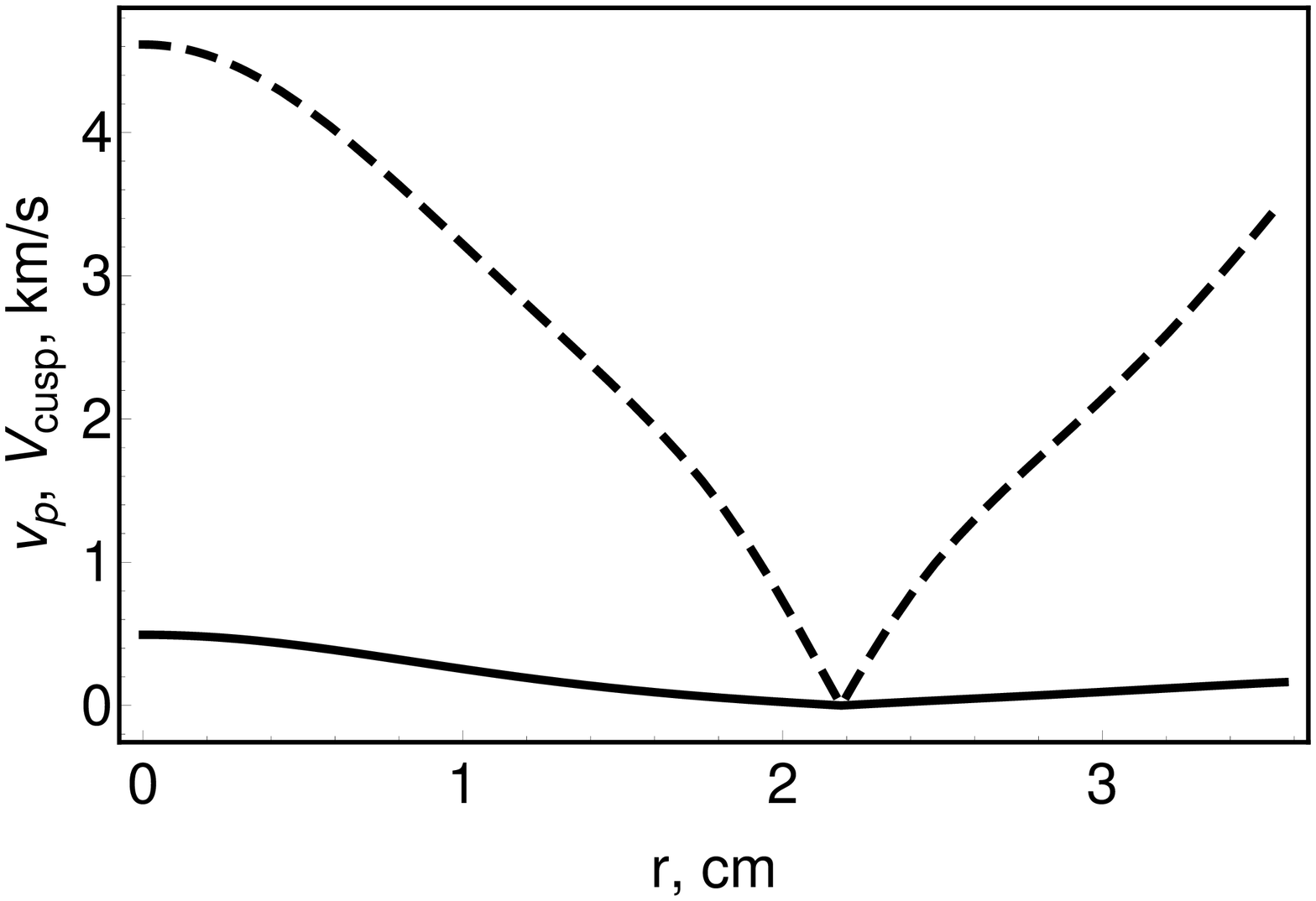} } \\ {\small b)}
	\end{minipage}
	\hfill
	\begin{minipage}{0.32\linewidth}
		\center{\includegraphics[width=1\linewidth]{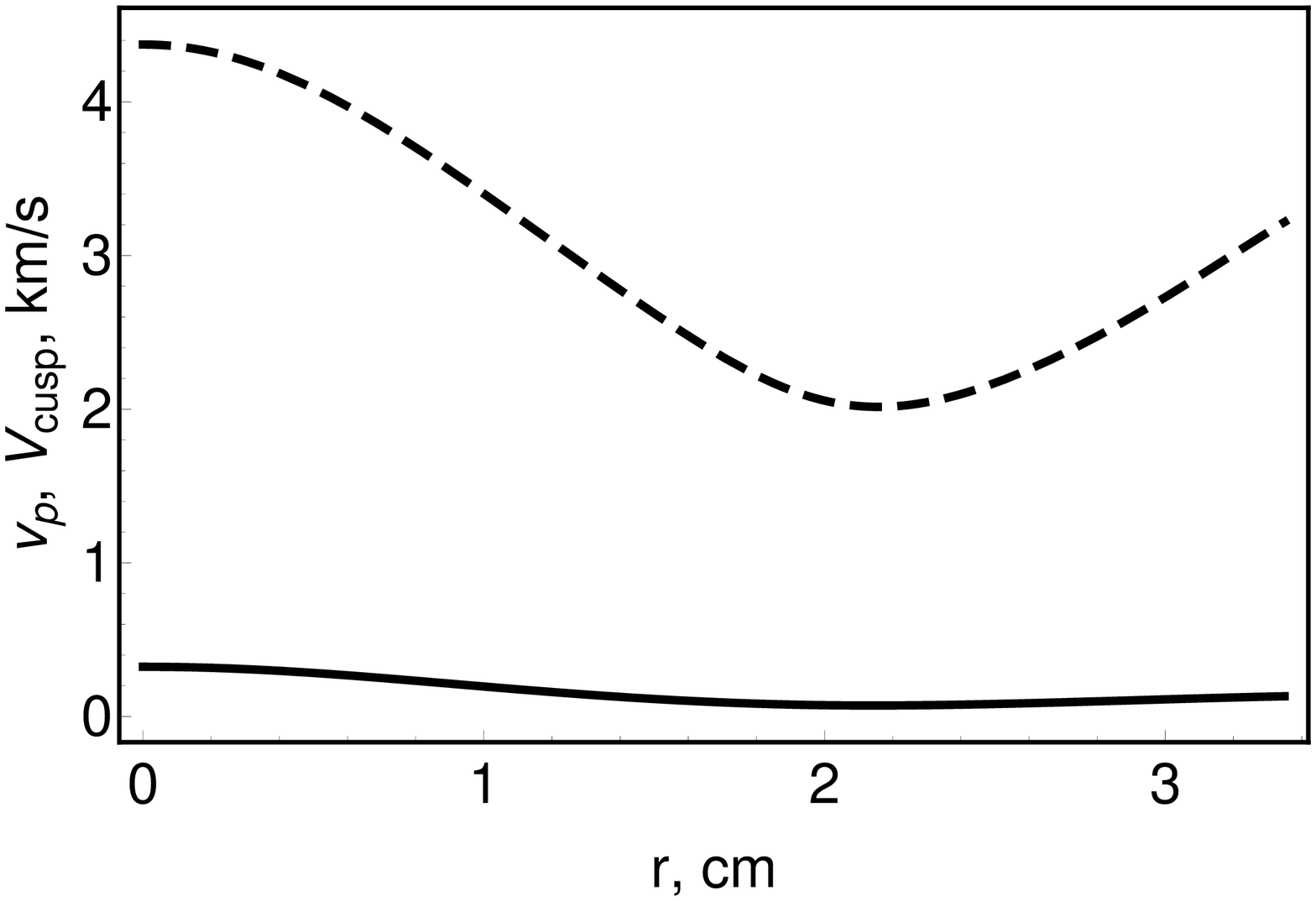} } \\ {\small c)}
	\end{minipage}
	\caption{Comparison of the poloidal, $v_{\rm p}$ (solid line), and cusp, 
	$V_{\rm cusp}$ (dashed line), velocities at various heights: (a) $z = 2$ cm, (b)
$z = -0.55$ cm, and (c) $z = -2$ cm.}
\label{cusp}	
\end{figure*}

Next, Fig.~\ref{density} shows the radial distributions of the number density, 
pressure, and temperature at various heights. Finally, Figs.~\ref{Mach} and 
\ref{cusp} (also for the three sections) show the square of the Alfv{\' e}n 
Mach number ${\cal M}^2$ and compare the poloidal velocity $v_{\rm p}$ and 
the cusp velocity $V_{\rm cusp}$ (\ref{Vcusp}). As we see, the conditions
${\cal M}^2 \ll 1$ and $v_{\rm p} \ll V_{\rm cusp}$  are actually fulfilled 
with a large margin.

\begin{center}
CONCLUSIONS
\end{center}
Thus, we found a new wide class of solutions for the generalized Grad-Shafranov 
equation that allows axisymmetric stationary subsonic flows to be described. 
Based on it, we determined the internal structure of the plasma outflow observed 
in the laboratory simulations of nonrelativistic jets at the KPF-4 ''Phoenix'' 
facility. 
 
Of course, it should be emphasized that the internal structure of the 
plasma jet was found only in some limited class of solutions for the 
Grad-Shafranov equation. Therefore, the approach considered here does 
not purport to be universal. On the other hand, the very fact that the 
Grad-Shafranov equation is linearized on a fairly wide class of integrals 
of motion (one free function $\eta_{\rm n}(\Psi)$  and four constants $A$, 
$\Omega_{0}$, $E_{0}$ and $K_{0}$) may already be considered as an 
independent important result of our work.

At the same time, it turned out that even such a simplified model well 
reproduces the main morphological characteristics of the plasma outflow 
--- the increase in its width in the tail and the presence of a characteristic
''funnel'' in the head. Accordingly, the presence of a narrow current channel 
near the jet axis (see Fig.~\ref{fields}) was also naturally explained. In future, 
it would be very useful to check the spatial distribution and other parameters 
(velocity, density, temperature) that are currently still inaccessible to 
direct measurements. The conclusion that a plasma circulation should inevitably 
emerge within the plasma outflow may be considered to be yet another interesting 
result of our analysis.

Finally, it should be emphasized once again that the solution obtained can 
be used as an initial condition when simulating the jet propagation at plasma
focus facilities. An analogous method could possibly also be applied to find 
self-consistent configurations of Herbig-Haro objects and then to numerically
compute their motion in the ambient medium.

As regards the astrophysical applications, it should be immediately noted 
here that the solution constructed above may be considered only as the first 
approximation. The point is that neither the dissipative heating processes 
nor the radiation processes, which play a prominent role in astrophysical 
sources, can be consistently described in terms of ideal magnetohydrodynamics.
Nevertheless, even such a simple model allowed the main morphological properties 
of flows, including the characteristic funnel in the jet head, to be reproduced. 
Naturally, a detailed analysis of all these questions was beyond the scope of this
paper.
 
\begin{center}
ACKNOWLEDGEMENTS    
\end{center} 
\vspace{0.2cm}
We are grateful to K.P. Zybin and V.I. Krauz for the stimulating discussion.

\begin{center}
FUNDING   
\end{center}  
\vspace{0.2cm}
 This work was supported by the Russian Foundation for Basic Research (project no. 18-29-21006).
 \vspace{0.2cm}
 
\begin{center}
REFERENCES 
\end{center} 
\begin{enumerate}
\item 
B. Albertazzi, A. Ciardi, M. Nakatsutsumi, et al., Science (Washington, DC, U. S.) 
{\bf 346}, 325 (2014).
\item  
H. G. Arce, D. Shepherd, F. Gueth, C.-F. Lee, R. Bachiller, A. Rosen, and H. Beuther, 
in {\it Protostars and Planets V}, Ed. by B. Reipurth, D. Jewitt, and K. Keil (Univ. 
of Arizona Press, Tucson, 2007), p. 245.
\item  
C. V. Atanasiu, S. G{\" u}nter, K. Lackner, and I. G. Miron, Phys. Plasmas {\bf 11}, 3510 (2004).
\item  
P. B. Bellan, J. Plasma Phys. {\bf 84}, 755840501 (2018)
\item  
V. S. Belyaev, G. S. Bisnovatyi-Kogan, A. I. Gromov, B. V. Zagreev, A. V. Lobanov, 
A. P. Matafonov, S. G. Moiseenko, and O. D. Toropina, Astron. Rep. {\bf 62}, 162 (2018).
\item  
V. S. Beskin, {\it Axisymmetric Stationary Flows in Astrophysics} (Fizmatlit, Moscow, 2005; 
Springrer, Heidelberg, 2010).
\item 
R. D. Blandford and D. G. Payne, Mon. Not. R. Astron. Soc. {\bf 199}, 883 (1992).
\item  
J. M. Blondin, B. A. Fryxell, and A. Ko¨ nigl, Astrophys. J. {\bf 360}, 370 (1990).
\item  
M. Bocchi, B.Ummels, J. P. Chittenden, S. V. Lebedev, A. Frank, and E. G. Blackman, Astrophys.
J. {\bf 767}, 84 (2013).
\item  
P. H. Bodenheimer, {\it Principles of Star Formation} (Springer, Heidelberg, 2011).
\item  
A. Ciardi, in {\it Jets from Young Stars IV}, Ed. by P. J. Valente Garcia, and J. M. Ferreira, 
Lect. Notes Phys. {\bf 793}, 31 (2010).
\item  
A. Ciardi, S. V. Lebedev, A. Frank, F. Suzuki-Vidal, G. N. Hall, S. N. Bland, A. Harvey-Thompson,
E. G. Blackman, et al., Astrophys. J. {\bf 691}, L147
(2009).
\item  
V. Duez and S. Mathis, Astron. Astrophys. {\bf 517}, A58 (2010).
\item  
A. Frank, T. P. Ray, S. Cabrit, P.Hartigan, H.G. Arce, F. Bacciotti, J. Bally, M. Benisty, 
J. Eisl{\" o}ffel, M. G{\" u}udel, S. Lebedev, B. Nisini, and A. Raga, in {\it Protostars 
and Planets VI}, (Ed. by H. Beuther, R. S. Klessen, C. P. Dullemond, and Th. Henning
(Univ. Arizona Press, Tucson, 2014), p. 451.
\item  
H. Grad, Rev. Mod. Phys. {\bf 32}, 830 (1960).
\item  
L. Guazzotto and E. Harmeiri, Phys. Plasmas {\bf 21} 022512 (2014).
\item  
E. C. Hansen, A. Frank, P. Hartigan, and S. V. Lebedev, Astrophys. J. {\bf 837}, 143 (2017).
\item  
G. Haro, Astron. J. {\bf 55}, 72 (1950).
\item  
G. H. Herbig, Astrophys. J. {\bf 111}, 11 (1950).
\item  
J. Heyvaerts and J. Norman, Astrophys. J. {\bf 347}, 1055 (1989).
\item  
M. Huarte-Espinosa, A. Frank, E. G. Blackman, A. Ciardi, P. Hartigan, S. V. Lebedev, 
and J. P. Chittenden, Astrophys. J. {\bf 757}, 66 (2012).
\item
P. Kajdi{\u c} and A. C. Raga, Astrophys. J. 670, 1173 (2007).
\item
V. I. Krauz, V. S. Beskin, and E. P. Velikhov, Int. J. Mod. Phys. D {\bf 27}, 1844009 (2018).
\item
V. I. Krauz, K. N. Mitrofanov, D. A. Voitenko, G. I. Astapenko, A. I. Markoliya, and 
A. P. Timoshenko, Astron. Rep. {\bf 63}, 146 (2019).
\item
V. Krauz, V. Myalton, V. Vinogradov, E. Velikhov, S. Ananyev, S. Dan’ko, Yu. Kalinin, 
A. Kharrasov, K. Mitrofanov, and Yu. Vinogradova, in {\it Proceedings of the 42nd EPS 
Conference on Plasma Physics} (2015), Vol. 39E, p. 4.401.
\item
V. I. Krauz, V. V. Myalton, V. P. Vinogradov, and E. P. Velikhov, J. Phys.: Conf. Ser. {\bf 907}, 012026 (2017).
\item
L. D. Landau and E.M. Lifshitz, {\it Course of Theoretical Physics, Vol. 8: Electrodynamics of 
Continuous Media} (Nauka,Moscow, 1982; Pergamon, New York, 1984).
\item
L. L. Lao, S. P. Hirshman, and R. M. Wieland, Phys. Fluids {\bf 24}, 1431 (1981).
\item
E. S. Lavine and S. You, Phys. Rev. Lett. {\bf 123}, 145002
(2019).
\item
S. V. Lebedev, A. Frank, and D. D. Ryutov, Rev. Mod. Phys. {\bf 91}, 025002 (2019).
\item
O. E. Lopez and L. Guazzotto, Phys. Plasmas {\bf 24}, 032501 (2017).
\item
Ch. F. McKee and E. C. Ostriker, Ann. Rev. Astron. Astrophys. {\bf 45}, 565 (2007).
\item
K. N. Mitrofanov, V. I. Krauz, V. V. Myalton, V. P. Vinogradov, A. M. Kharrasov, 
and Yu. V. Vinogradova, Astron. Rep. {\bf 61}, 138 (2017).
\item
M. L. Norman, K.-H. Winkler, L. Smarr, and M. D. Smith, Astron. Astrophys. {\bf 113}, 285 (1982).
\item
G. Pelletier and R. E. Pudritz, Astrophys. J. {\bf 394}, 117 (1992).
\item
A. C. Raga, F. de Colle, P. Kajdi{\u c}, A. Esquivel, and J. Cant{\' o}, Astron. Astrophys. 
{\bf 465}, 879 (2007).
\item
T. Ray, C. Dougados, F. Bacciotti, J. Eisl{\" o}ffel, and A. Chrysostomou, in {\it Protostars 
and Planets V}, Ed. by B. Reipurth, D. Jewitt, and K. Keil (Univ. Arizona Press, Tucson, 2007), 
p. 231.
\item
B. Reipurth, S. Heathcote, J.Morse, P. Hartigan, and J. Bally, Astron. J. {\bf 123}, 362 (2002).
\item
D.D. Ryutov, M. S. Derzon, and M. K. Matzen, Rev. Mod. Phys. {\bf 72}, 167 (2000).
\item
V. D. Shafranov, Sov. Phys. JETP {\bf 6}, 545 (1957).
\item
L. S. Soloviev, in {\it Reviews of Plasma Physics}, Ed. by M. A. Leontovich 
(Atomizdat, Moscow, 1963), Vol. 3, p. 245;  (Consultants Bureau, New York, 1965), 
Vol. 3, p.  28.
\item
B. U. {\~ A} . Sonnerup, H. Hasegawa, W.-L. Teh, and L.-N. Hau, J. Geophys. Res. 
{\bf 111}, A09204 (2004).
\item
J. M. Stone and M. L. Norman, Astrophys. J. {\bf 413}, 210 (1993).
\item
J. M. Stone and Ph. E. Hardee, Astrophys. J. {\bf 540}, 192 (2000).
\item
V. G. Surdin, Birth of Stars (URSS,Moscow, 2001) [in Russian].
\item
F. Suzuki-Vidal, M. Bocchi, S. V. Lebedev, G. F. Swadling, G. Burdiak, S. N. Bland,
P. de Grouchy, G. N. Hall, et al., Phys. Plasmas {\bf 19}, 022708 (2012).
\item
O. Te{\c s}ileanu, A. Mignone, S. Massaglia, and F. Bacciotti, Astrophys. J. {\bf 746}, 96 (2012).
\item
Ya. B. Zel’dovich, S. I. Blinnikov, and N. I. Shakura, {\it Physical Foundations of the 
Structure and Evolution of Stars} (Mosk. Gos. Univ., Moscow, 1981) [in Russian].
\end{enumerate}

\end{document}